\documentclass[a4paper]{EslabStyle}
\usepackage{epsfig}

\def\bdm{\begin{displaymath}} 
\def\edm{\end{displaymath}} 
\def\beq{\begin{equation}} 
\def\eeq{\end{equation}} 
\def\bit{\begin{itemize}} 
\def\eit{\end{itemize}} 
\def\ben{\begin{enumerate}} 
\def\een{\end{enumerate}} 
\def\bfi{\begin{figure}[htb]} 
\def\bpfi{\begin{figure}[p]}
\def\mum{$\>{\mu {\rm m}}$}

\def\lsol{$\rm L_{\odot}$}
\def\msol{$\rm M_{\odot}$}

\def\mtol{$M/L$}
\def\mtolk{$M/L_K$}

\def\ea{{\it et al.~}}

\newcommand{\kms}{\>{\rm km}\,{\rm s}^{-1}}

\newcommand{\Msun}{\>{\rm M_{\odot}}}

\begin{document}

\title{Star formation in the nuclei of spiral galaxies}

\author{T. B\"oker\inst{1,2}} \institute{Space Telescope Science Institute
-- 3700 San Martin Drive -- Baltimore, MD 21218 -- U.S.A. \and
Astrophysics Division -- Space Science
  Department of ESA, ESTEC, Postbus 299, NL-2200 AG Noordwijk, The
  Netherlands }

\maketitle 
\begin{abstract}
Recent observations with the Hubble Space Telescope (HST) 
have revealed that a large fraction of late-type (Sc and later) spiral galaxies
harbor a bright, compact stellar cluster in their dynamical centers. 
Statistics of the mass, age, and star formation history of these clusters 
as a function of their host galaxy's Hubble type
can be used to constrain models of secular galaxy evolution.
Since late-type spirals by definition do not possess a prominent bulge,
their nuclear clusters are more easily separated from the underlying
disk population. Their spectroscopic properties can thus be 
studied from ground-based observations. 
Here, I will discuss plans for, and first results of, a 
program to study a sample of known nuclear clusters in 
late-type spirals. For one galaxy (IC~342), we have used 
high-resolution near infrared spectroscopy to determine the cluster mass 
directly via its stellar velocity dispersion. 
The analysis conclusively shows a very low mass-to-light ratio for
the nuclear cluster in \object{IC~342},
indicative of a young cluster age ($\approx$ 50~Myrs). From probability
arguments, this result favors the scenario that 
such bursts are a recurrent phenomenon in late-type spiral nuclei.
\end{abstract}
\section{Introduction}
Galactic nuclei are a unique physical environment as evidenced by
massive black holes, active galactic nuclei of various flavors,
and a star formation history that is very different from their
host disks. While it is plausible that the nucleus might have unusual 
properties in galaxies with steep, cuspy light profiles, the gravitational
potential vanishes in diffuse, non-singular galaxies without a central black 
hole. In these cases it not at all intuitive why the nucleus should be
the site of vigourous star formation.
It therefore came as a surprise when recent HST data revealed the presence
of prominent compact nuclear sources in many spiral galaxies of all
Hubble types (\cite{phi96,mat99,boe99a,car99}). More specifically, this is true for
virtually all galaxies with exponential bulges - mostly galaxies of type
Sc or later. Because the nuclear sources in most cases
are resolved with HST, they are interpreted as stellar clusters.

The processes that lead to the formation of stellar clusters 
in the very center of galaxies are not well understood, and many
questions remain to be answered:  
Is the cluster formation process self-regulating in the sense all clusters
form over similar timescales? How stable are nuclear clusters? 
Is a stellar bar required to funnel gas into the central regions?

The presence of such clusters is also interesting in the context of
the ongoing debate on galaxy bulge formation.
It is known from numerical simulations
(e.g. \cite{nor96}) that the growth of a central mass concentration can
destroy stellar bars on short timescales. The bar undergoes a vertical 
"buckling" instability, during which stars are heated into orbits
above the galaxy plane. The result is a triaxial system that much 
resembles a galaxy bulge. 

While this process is both plausible and appealing, there are also problems 
related to it. First, the observed high fraction of barred galaxies seems to 
contradict the notion that stellar bars are
easily destroyed by central mass concentrations. Second, if
the above scenario was true, then destruction of the bar would disable an
effective funneling mechanism for the gas to reach the galactic center. As
a consequence, repeated nuclear cluster formation is prevented. 

Both these arguments can in principle be overcome if more and more 
massive bars form in a recurrent way. Building a galaxy bulge via repeated 
cycles of the bar formation - gas inflow - cluster formation - bar 
destruction cycle therefore seems like a possible alternative
to theories in which the bulge properties are largely determined during 
the collapse of the primordial gas cloud.
\begin{figure}[ht]
  \begin{center}
    \epsfig{file=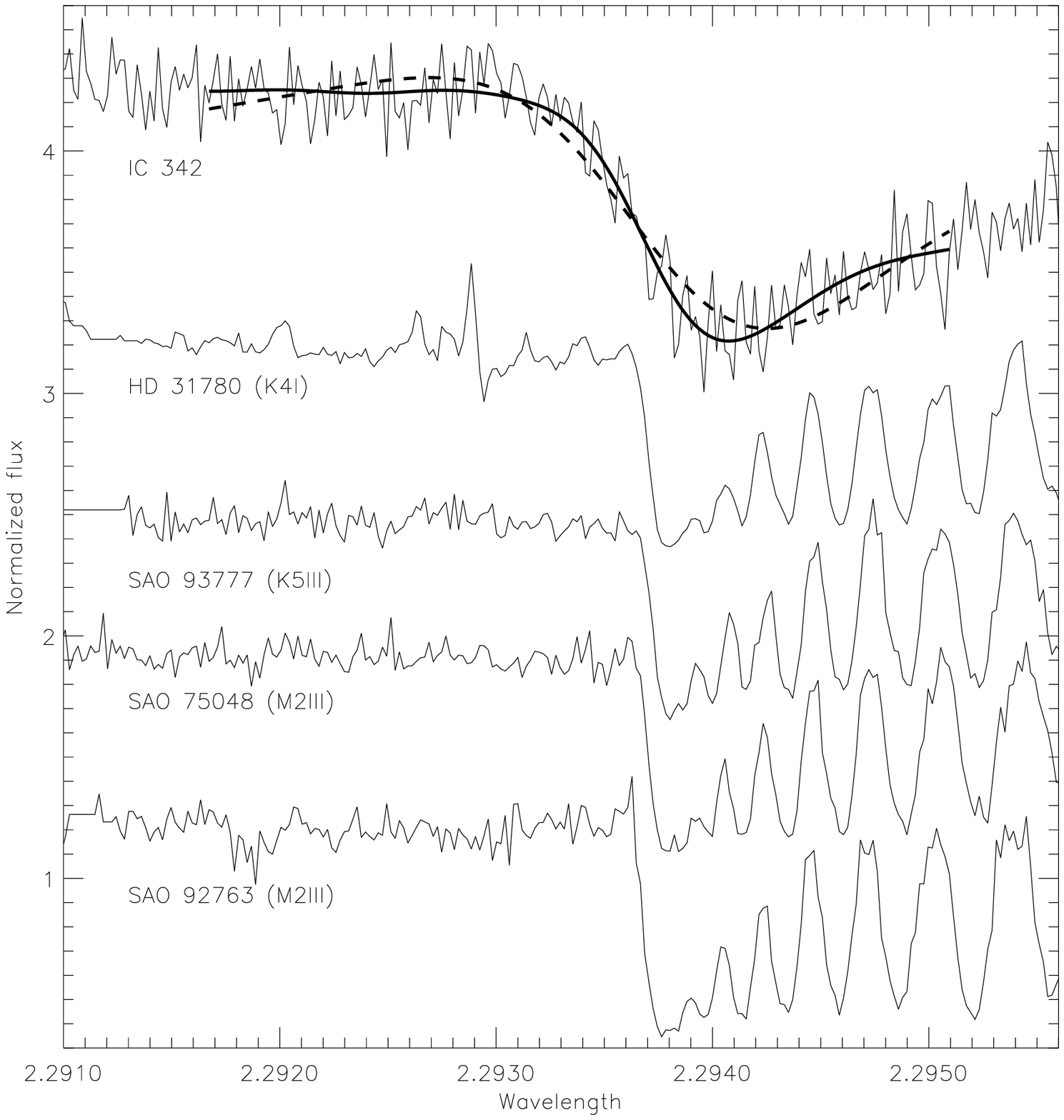, width=9cm}
  \end{center}
\caption{Normalized CO-bandhead spectrum
of the nuclear cluster in IC~342, compared to four late-type template
stars. All spectra are offset vertically for clarity; 
the abscissa is in \mum . The heavy
solid curve shows the best Gaussian-convolved template fit to the
galaxy spectrum, which has $\sigma = 33 \kms$. By contrast, the heavy
dashed curve shows the (unacceptable) fit for a fixed value of
$\sigma = 62 \kms$. The template spectra themselves show the 
instrumental dispersion of $\sigma = 5.5 \kms$. \label{fig:spec}}
\end{figure}

Understanding the formation history of nuclear star clusters provides
a new and important diagnostic to test the above scenario. 
Reliable statistics of nuclear cluster ages will directly test whether
or not they form in a repetitive cycle. For this reason, we have initiated a
program to obtain high-quality spectra of the nuclear clusters of
a sample of late-type, face-on spirals for which the cluster can be
separated from the galaxy disk in ground-based observations. 

In this paper,  
I will describe our methods to derive the cluster ages from the observations
and give an example for the successful application in Sec.~\ref{sec:dating}. 
Sec.~\ref{sec:formation} discusses the preliminary evidence for 
repetitive cluster formation, and Sec.~\ref{sec:plans} 
describes plans for an observational program
which will put the analysis on much firmer statistical footing. 
\section{Age-dating nuclar clusters}
\label{sec:dating}
\begin{figure}[ht]
  \begin{center}
    \epsfig{file=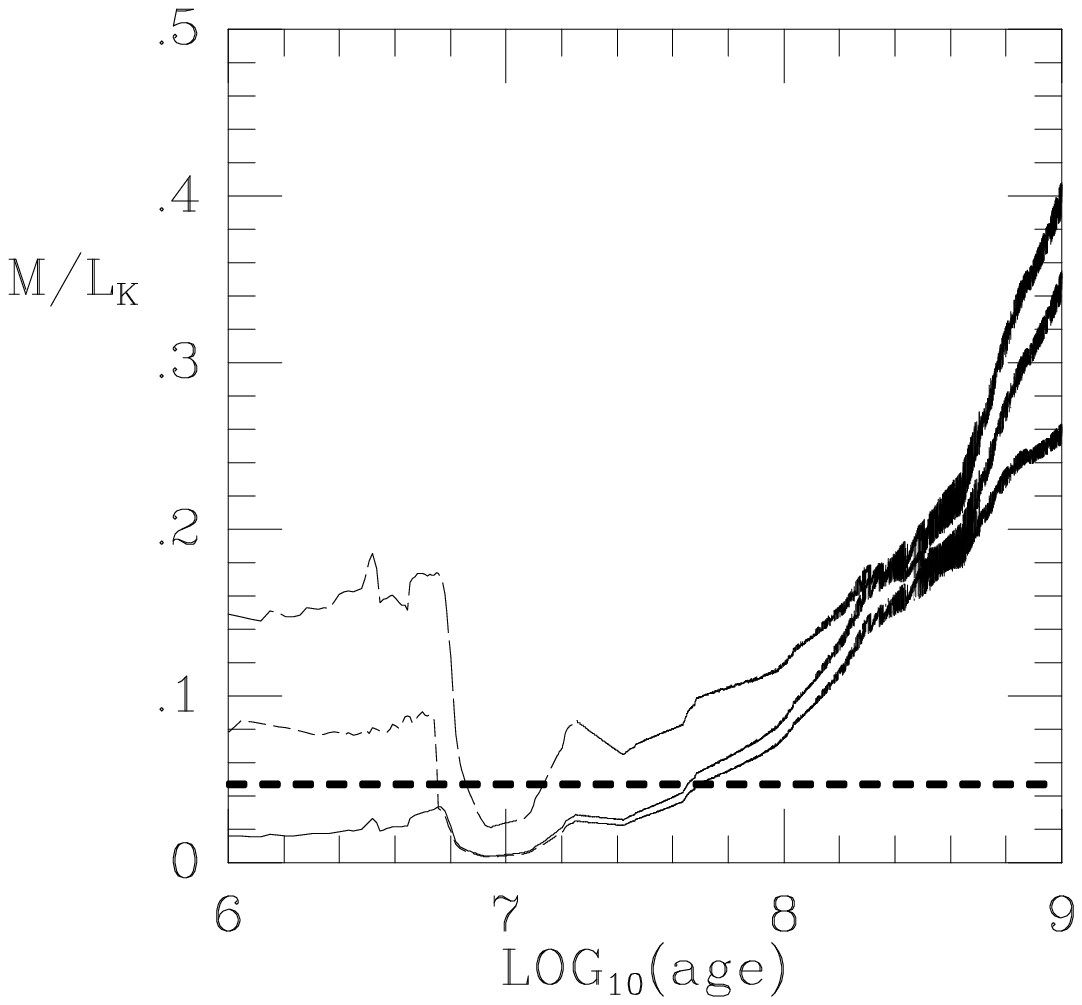, width=9cm}
  \end{center}
\caption{Relation between $M/L_K$ (in solar units) and age (in years) 
for several model stellar clusters (from Leitherer \ea 1999).
Different curves are models that
differ in the slope $\alpha$ and upper-mass cut-off $M_{\rm up}$ of
their IMF. Solid: $\alpha = 2.35$ and $M_{\rm up} = 100 \Msun$;
long-dashed: $\alpha = 3.30$ and $M_{\rm up} = 100 \Msun$;
short-dashed: $\alpha = 2.35$ and $M_{\rm up} = 30 \Msun$.  All
models assume solar metallicity  ($Z = 0.02$), a lower-mass IMF 
cut-off at $M_{\rm low} = 1 \Msun$, and an instantaneous burst. 
The dashed horizontal line indicates the value $M/L_K = 0.047$
inferred for the nuclear star cluster of IC 342.
The models indicate that the cluster has $\log({\rm age}) \leq 7.7$. 
\label{fig:models} }
\end{figure}
It is well known from a number of population synthesis codes 
(e.g. \cite{bru93,lei99}) that the mass-to-light ratio \mtol\ of
a stellar cluster is a sensitive age indicator. We have chosen the
K-band for our analysis for the following reasons. Firstly, galactic
nuclei are often affected by high and non-uniform dust obscuration.
The near-infrared is much less sensitive to extinction than the optical:
$A_K\approx 0.1\>A_V$. Secondly, the near-infrared emission is dominated by
the evolved stellar population, and therefore gives a better representation
of the stellar mass distribution than optical bands in which 
gaseous emission lines often contribute significantly. For both reasons,
it is often easier to identify the true nucleus of a galaxy at near-infrared
wavelengths. The third reason for choosing K-band is that it contains
the $\rm ^{12}CO\>(0-2)$ bandhead at 2.2936\mum , a prominent absorption
feature which arises in the atmospheres of cool giants and supergiants.
This feature is located in a spectral region with favorable observing
conditions and has been shown to be a powerful diagnostic for stellar 
kinematics in external galaxies, as discussed in detail by \cite*{gaf95}.
\begin{figure*}[ht]
  \begin{center}
    \epsfig{file=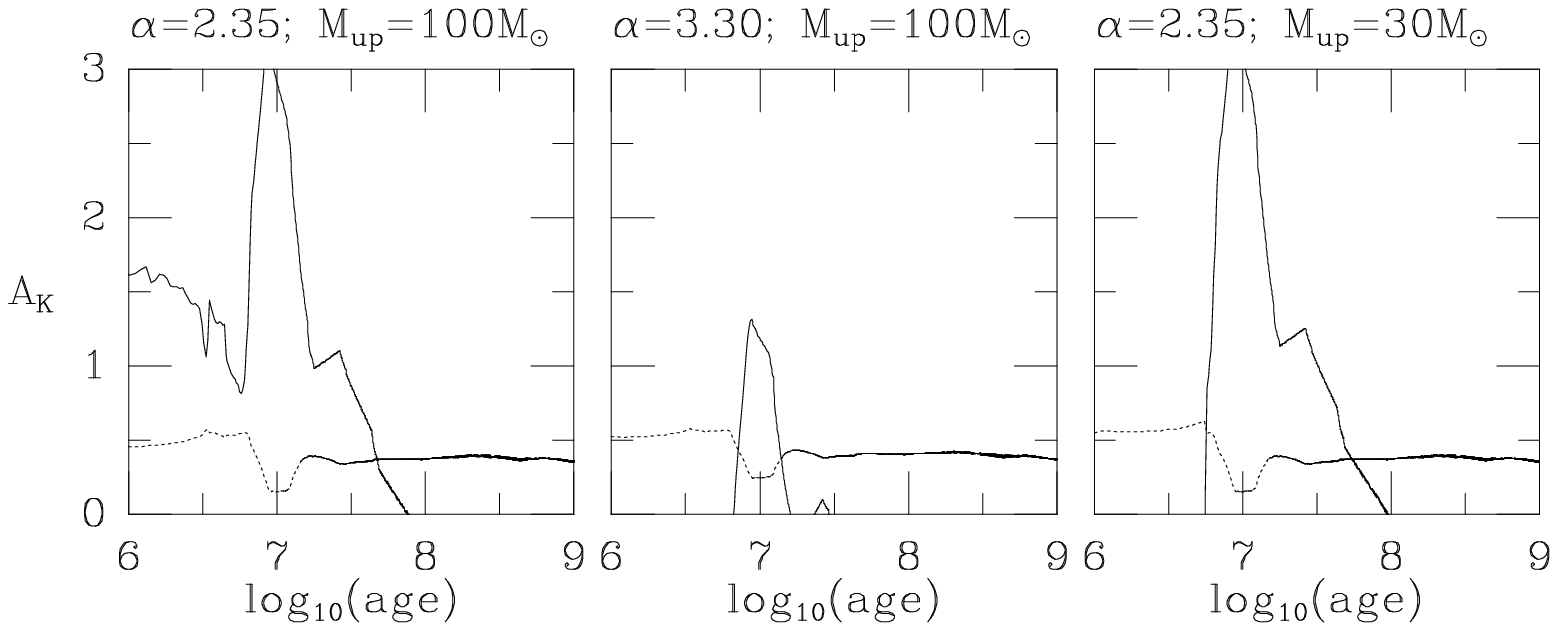, width=18cm}
  \end{center}
\caption{Illustration of how both age (in years) 
and $A_K$ of the nuclear cluster in IC~342 can be determined 
from the observed $(V-K)_{\rm
obs}=4.57$ and $M/L_K = 0.047 \times 10^{-0.4 (A_K - 0.45)}$. Each
panel corresponds to one of the three population models in
Fig.~\ref{fig:models}. The dotted curves
show the $A_K$ for which the predicted $V-K$ of the population model
matches the observations, and the solid curves show the $A_K$ for
which the predicted \mtolk\ matches the observations. The
intersections of the two curves
indicate combinations of age and $A_K$ that fit both observational
constraints. The allowed combinations agree with the assumed extinction 
and the derived age, demonstrating self-consistency. \label{fig:ext} }
\end{figure*}

Our methodology for deriving the \mtol\ ratio is based on the ``Gauss-Hermite
Pixel Fitting Software'' developed by \cite*{mar94}. In brief, 
the observed surface brightness profile is
deprojected under the assumption of spherical symmetry, and the
three-dimensional velocity dispersion profile $\sigma(r)$ is
calculated by solving the Jeans equation for a spherical isotropic
system. The results are then projected along the line of sight and
convolved with the instrumental profile to yield a prediction for
the observed velocity dispersion. This dispersion scales as $\sigma
\propto \sqrt{M/L}$, and we determine the \mtol\ that produces the
observed $\sigma$.

Two observational ingredients are needed for the above recipe to
yield a reliable \mtol\ value. In addition to a high-quality, high-resolution
spectrum - yielding the cluster's velocity dispersion $\sigma$ - it is
essential to also obtain an accurate surface brightness profile. Typical 
diameters of nuclear clusters are 0.2\arcsec\ (\cite{car99}), therefore
imaging with HST-like resolution is required. 
\subsection{The showcase: IC~342} \label{sec:ic342}
We recently were able to successfully apply the above method
to the nuclear cluster of IC~342 (\cite{boe99b}). For these
observations, we used the near-infrared spectrograph CSHELL 
(\cite{tok90,gre93}) at the NASA InfraRed Telescope Facility (IRTF) to
determine the $\sigma$ from the shape of the
CO~(0-2) bandhead. Figure~\ref{fig:spec} shows the
spectrum in comparison to a number of stellar templates that were used
to derive a value of $\sigma = 33\pm 3\kms$.
Together with an archival HST V-band image, we used the observed 
velocity dispersion to derive the cluster mass-to-light ratio according 
to the method outlined in \ref{sec:dating}. Our data imply
\mtol = 0.05~\msol /\lsol\ and a total cluster mass of $6\cdot 10^6\Msun$.

In order to derive the cluster age, we compared these results to predictions
from population synthesis models such as the ``Starburst99'' package of
\cite*{lei99}. We used an instantaneous burst model for our comparison
which is justified by the high equivalent width (EW) of the CO absorption
as demonstrated in an earlier paper (\cite{boe97}, hereafter BFG97).
Fig.~\ref{fig:models} shows that - quite independent of the
assumed Initial Mass Function (IMF) - the nuclear cluster in IC~342 is
rather young, namely less than 50~Myrs.
\subsection{Dealing with extinction}
In galactic nuclei, dust extinction can significantly affect the
observations, even at nearinfrared wavelengths. However, if values
for both color and mass-to-light ratio are available from the
observations, it is possible to perform a consistency check on the
assumed extinction. This is demonstrated in Fig.~\ref{fig:ext} which
for three different IMF's shows possible combinations of age and 
extinction which explain the observed $(V-K)$ color and
\mtolk\ ratio, respectively. Locations where the two lines intersect indicate 
age/extinction combination which are consistent with both observables. 
The allowed combinations have ages in
the range $10^{6.8-7.8}$ years, and $A_K$ between
$0.37$ and $0.57$. This range includes the value $A_K \approx 0.45$
(i.e., $A_V = 4.0$) that was adopted for our analysis based on the
results of BFG97, and the age that we derived for the nuclear cluster.
\section{Repetitive cluster formation?}
\label{sec:formation}
Given the young age and low mass-to-light ratio of the IC~342 nuclear cluster, 
one is led to ask whether its formation was the first of
this kind. The presence of an underlying older population 
is not ruled out by our observations, but there are
constraints on the amount of light and mass that such a population
could contribute. These constraints are 
based on the observed EW of the CO bandhead at 2.29\mum\ which is given by 
${\rm EW}_{\rm total} = {\rm EW}_{\rm young} - f({\rm EW}_{\rm young} - {\rm
EW}_{\rm old})$. Here, $f$ is the fraction of the total light
contributed by an old population. The EW for single-burst population
models decreases with age, so the
observed EW sets an upper limit on $f$, namely $f \leq 0.14$. 

Although an old population can only contribute a small fraction of the 
observed light, its contribution to the total mass of the cluster may be
significant. For the same mass, a burst that happened $10^{9.5}$~yrs
ago would be 3 magnitudes fainter in the K-band than the young
population that dominates the light (\cite{bru93}). 
An old population that formed $10^{10.2}$~yrs ago,
roughly the age of the universe, would be 5 magnitudes fainter.
In the latter case it could be $15$ times as massive as the observed 
young population and still only have $f\approx 0.14$. The total mass 
of the cluster is fixed, so in this scenario the
young population would need to have a mass-to-light ratio that is
15 times smaller than previously inferred, which implies
$M/L_K \approx 0.003$. This is roughly the smallest value that can be
plausibly explained with a single-burst population model
(cf.~Fig.~\ref{fig:models}), and cannot be ruled out. So while we
can conclude that the light of the cluster is dominated by a young
population, this is not necessarily true for the mass.
\begin{figure}[ht]
  \begin{center}
    \epsfig{file=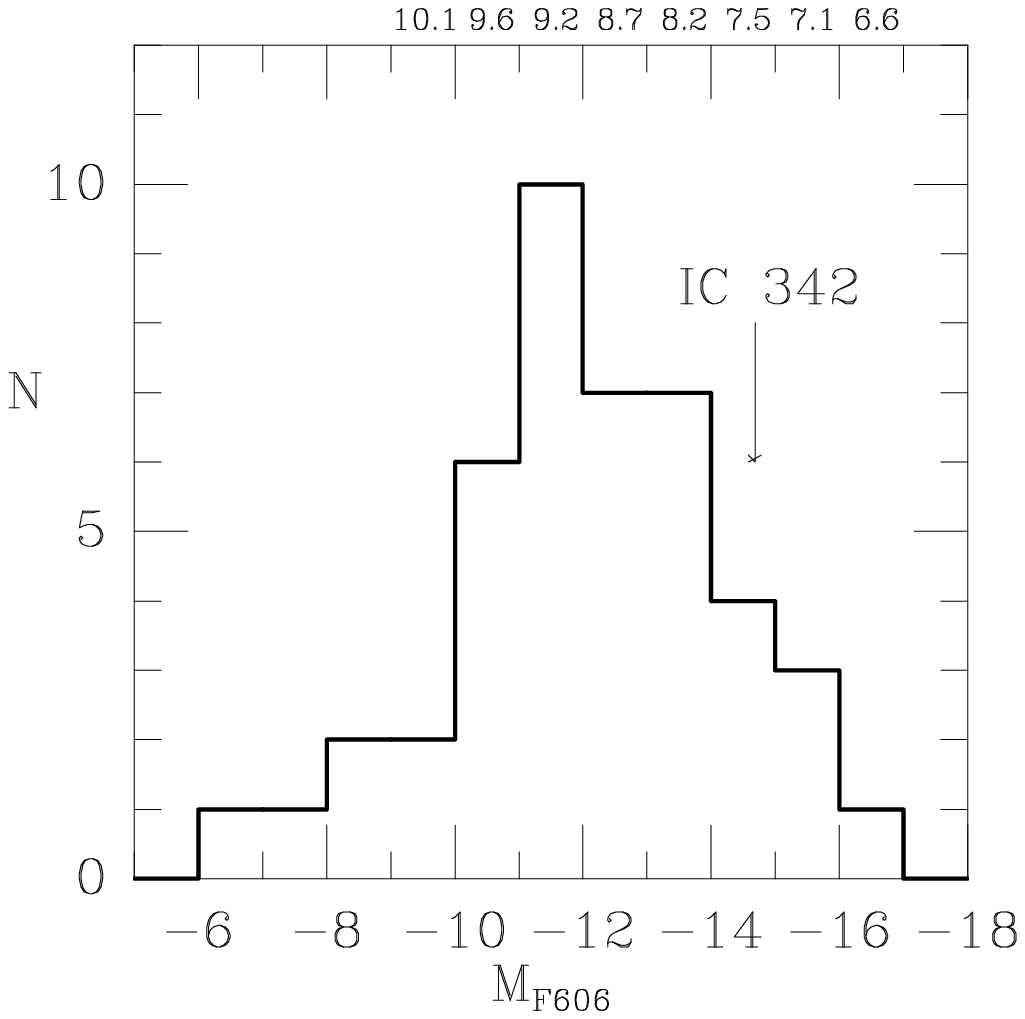, width=9cm}
  \end{center}
\caption{Histogram of nuclear stellar cluster 
luminosities for the galaxy sample of Carollo (1999).
Only galaxies with detected nuclear clusters are included. 
IC 342 is indicated by the
arrow, but is itself not included in the sample used to construct
the histogram. The labels at the top indicate $\log({\rm age})$ for
each bin, using a simple model that assumes that all clusters are
identical except for their age. The luminosities were calculated from
the stellar population models of Bruzual \& Charlot (1993), and are
normalized to $\log({\rm age}) = 7.4$ for IC 342. \label{fig:histo} }
\end{figure}

For the Galactic center, spectroscopy of individual stars is possible, 
and such data provide solid evidence for multiple periods of star formation
(\cite{kra95}).
In general, however, it is a difficult task to identify multiple star 
formation events from the integrated light of stellar populations, and even
more so to find a unique solution to their star formation history.
The best hope for answering whether or not stellar clusters form
repeatedly in the nuclei of spiral galaxies stems from a statistical analysis
of their frequency and age. 
\begin{figure*}[ht]
  \begin{center}
    \epsfig{file=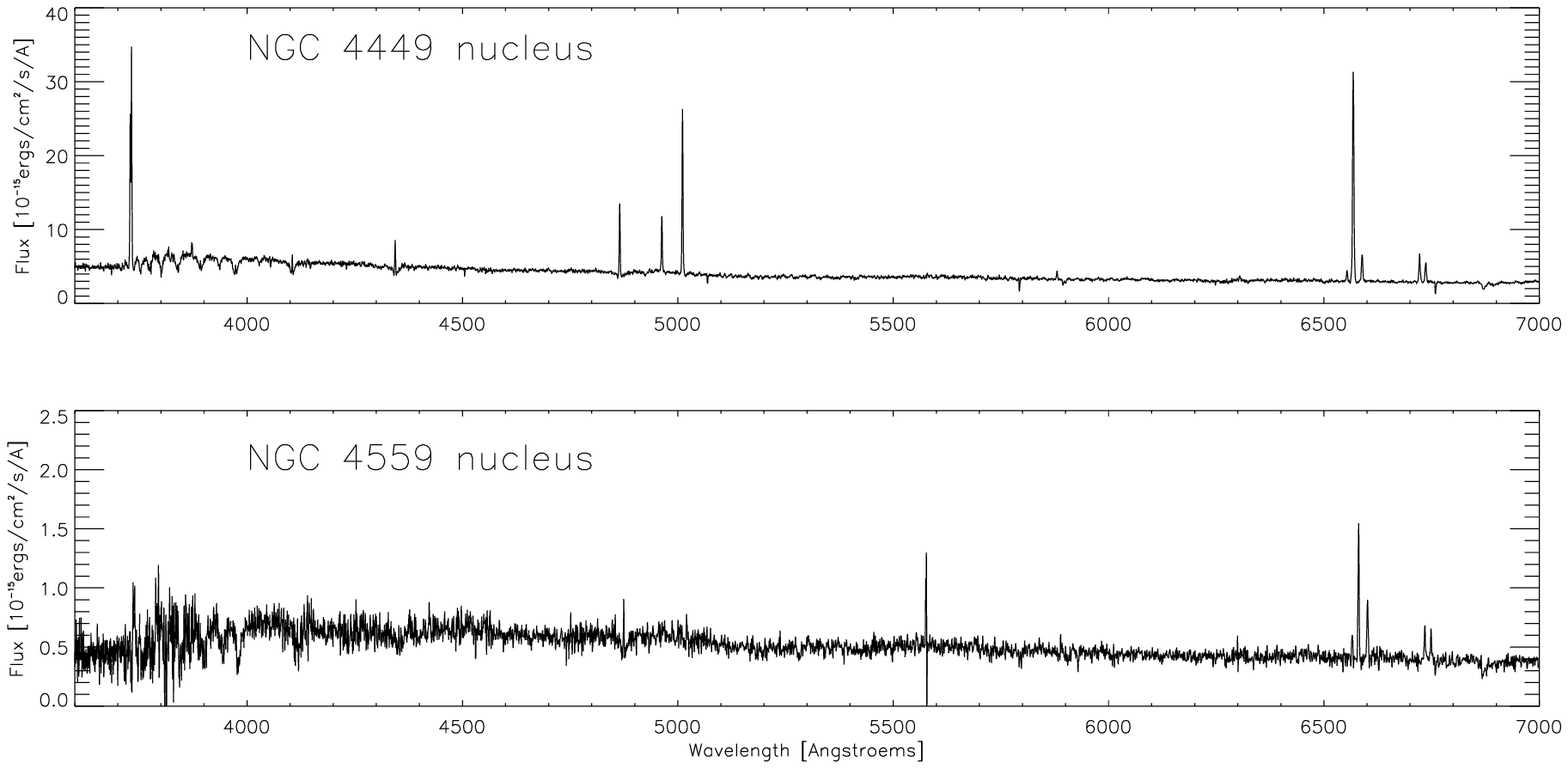, width=18cm}
  \end{center}
\vspace*{-9cm}
\caption{Echelle spectra of the nuclear clusters of NGC 4449
(top) and NGC 4559 (bottom). The spectrum of the underlying disk/bulge
population has been subtracted. The
feature at 5577~\AA\ in the spectrum of NGC~4559 is a residual skyline.
Both clusters show deep Balmer absorption features, indicative of 
a young cluster age.
\label{fig:echelle} }
\end{figure*}
\subsection{Statistics so far}
IC~342 is not the only galaxy known to harbor a young population of
stars in its nucleus. The Milky Way, for example, underwent a recent
starburst some 3 Myrs ago (\cite{kra95}), and both M31 and M33 have blue
nuclei that are quite possibly young star clusters (\cite{lau98}). If
nuclear cluster formation is a Poissonian process with equal probability
per unit time, one would expect to find clusters with ages in the range
of $10^{7\pm0.5}$, $10^{8\pm0.5}$, and $10^{9\pm0.5}$ years in the 
ratio $1:10:100$. 

In order to test this prediction in the absence of 
reliable age estimates for a large number of clusters, we took the following
approach. Under the very simplifying assumption that all nuclear clusters
have the same mass as the IC~342 cluster,
and formed in an instantaneous burst, one can estimate their ages
directly from luminosities. This is shown in Fig.~\ref{fig:histo} for
all clusters found in the \cite*{car99} sample.
While the interpretation of the Fig.~\ref{fig:histo} is certainly
hindered by the extremely simplifying assumption of equal mass and formation
history for all clusters, it nevertheless
seems that there is an overabundance of relatively young clusters which
can be interpreted as preliminary evidence for a repetitive process. 
\section{Observational plans}
\label{sec:plans}
In order to improve the available dataset on nuclear clusters, we are
planning the following observations:
\bit
\item{High-resolution CO bandhead spectroscopy (NIRSPEC @ Keck - approved)
of all galaxies found to date with HST to harbor a bright 
($K\leq$ 16) nuclear cluster with the goal to derive stellar velocity
dispersions.
}
\item{Optical Echelle spectroscopy 
(BC spectrograph @ Steward 90\arcsec\ - data taken)
of the same targets to constrain their stellar populations. 
Figure~\ref{fig:echelle} shows "quick-look" results of the
data reduction for two of the sample galaxies. 
In both cases, the (disk-subtracted) cluster spectra show deep
Balmer absorption lines indicative of fairly young cluster ages.
The detailed analysis of these spectra is ongoing, and will
be published separately (\cite{maz00}). 
}
\item{NIR imaging of about 100 late-type, face-on spirals (MAGIC @ MPIA 2.2~m 
- approved) in order to identify a larger sample of prominent nuclear cluster.
}
\item{A cycle~9 HST WFPC2 R-band snapshot survey (proposed) to identify a large
number of nuclear clusters for follow-up spectroscopy
{\bf and} obtain their surface brightness profiles.
}
\eit
The combination of ground-based spectroscopy and HST imaging 
will allow us to measure cluster masses and ages in
a way similar to that described for IC~342 for a large sample of galaxies.
The cluster mass and age distribution as a function of host galaxy Hubble type
will answer whether the formation of the nuclear
clusters in spiral galaxies is a recurring phenomenon or not, and how
- if at all - these clusters are related to the secular evolution of the
exponential bulges in which they reside.
\begin{acknowledgements}
It is a pleasure to acknowledge the contributions of my
collaborators R.P. van der Marel, W.D. Vacca, H.-W. Rix,
L. Ho, and J. Shields. I would also like to thank the 
LOC for organising a very lively and stimulating meeting.
\end{acknowledgements}
\end{document}